\documentstyle[aps,preprint,tighten,psfig]{revtex}
\newcommand{\lsim}{\mathrel{\mathop{\kern 0pt \rlap
  {\raise.2ex\hbox{$<$}}}
  \lower.9ex\hbox{\kern-.190em $\sim$}}}
\newcommand{\gsim}{\mathrel{\mathop{\kern 0pt \rlap
  {\raise.2ex\hbox{$>$}}}
  \lower.9ex\hbox{\kern-.190em $\sim$}}}
\newcommand{\beq}    {\begin{equation}}
\newcommand{\eeq}    {\end{equation}}
\newcommand{\beqarr} {\begin{eqnarray}}
\newcommand{\eeqarr} {\end{eqnarray}}
\newcommand{\barr}   {\begin{array}}
\newcommand{\earr}   {\end{array}}

\begin{document}

\preprint{
\begin{tabular}{r}
DFTT 51/2000
\end{tabular}
}

\title{Implications of a possible 115 GeV supersymmetric Higgs boson \\
  on detection and cosmological abundance of relic neutralinos}

\author{\bf 
A. Bottino, 
N. Fornengo and S. Scopel \footnote{E--mail: bottino@to.infn.it, 
fornengo@to.infn.it, scopel@to.infn.it}}

%\vspace{2cm}

\address{
\begin{tabular}{c}
Dipartimento di Fisica Teorica, Universit\`a di Torino \\
and INFN, Sez. di Torino, Via P. Giuria 1, I--10125 Torino, Italy\\
\end{tabular}
}

\maketitle

\begin{abstract}
  We show that a supersymmetric neutral Higgs boson with a mass of
  about 115 GeV and with the other prerequisites required by the LEP
  Higgs events would be compatible with the detection of relic
  neutralinos in current set--ups for WIMP direct search. Thus this
  putative Higgs would fit remarkably well in an interpretation in
  terms of relic neutralinos of the annual--modulation effect recently
  measured in a WIMP direct experiment. We also show that the
  cosmological abundance of the relevant neutralinos reaches values of
  cosmological interest.
\end{abstract}  

\vspace{1cm}


\section{Introduction}

Recent LEP data at center-of-mass energy above 206 GeV have provided a
hint for a Higgs boson with a mass of about 115 GeV \cite{leplast}.
The analysis presented in Ref.  \cite{leplast} refers to a Standard
Model (SM) Higgs. However, the very fact that the mass of this
putative neutral boson is relatively light entails the possibility for
this particle to be a neutral Higgs boson in the framework of a
supersymmetric extension of the SM; in fact a light Higgs is just what
one would expect in a susy scheme \cite{who}.

The existence of a susy neutral Higgs with a mass of about 115 GeV
would have important consequences for various aspects \cite{je}; in
the present paper we analyze its possible implications for dark
matter.  We prove that a susy neutral Higgs boson with a mass of 115
GeV and with the other prerequisites required by the LEP Higgs events
would be quite adequate to make scattering processes of relic
neutralinos off nuclei detectable in the current apparata for WIMP
direct search \cite{morales,damalast,cdms}. We recall that, taking
into account the present uncertainties in astrophysical quantities,
the sensitivity of the current experiments for WIMP direct
measurements, in the WIMP mass range: 40 GeV $\leq m_\chi \leq $ 200
GeV, may be established to be \cite{our}

\begin{equation}
4 \cdot 10^{-10} \; {\rm nbarn} \leq \xi \sigma^{\rm (nucleon)}_{\rm scalar} \leq 
 2 \cdot 10^{-8} \; {\rm nbarn}, 
\label{eq:section}
\end{equation}
\noindent
where $\sigma^{\rm (nucleon)}_{\rm scalar}$ is the scalar WIMP--nucleon
scattering cross section and $\xi \equiv \rho_\chi / \rho_l$ ($\rho_l$ is the
local value for the non--baryonic dark matter; we recall that the range for
$\rho_l$ is 0.2 GeV cm$^{-3} \leq \rho_l \leq$ 0.7 GeV cm$^{-3}$ \cite{t}).
This applies to WIMPs which interact with nuclei dominantly by coherent
effects, and with equal strength with neutrons and protons.  In this note we
show that the putative susy Higgs would fit remarkably well in an
interpretation of the annual--modulation effect measured in WIMP direct
searches \cite{damalast} in terms of relic neutralinos \cite{our,noi,an}. We
also derive that the cosmological abundance of the relevant neutralinos reach
values of cosmological interest.

Since the analysis of the recent LEP data in terms of susy Higgs
bosons by the LEP Collaborations is not available yet, we perform here
an independent, approximate estimate of the susy configurations which
would be involved in the LEP Higgs events. Our derivation makes use of
a number of simplifying assumptions, but, we believe, it is adequate
to outline the quite intriguing perspectives of the relevant scenario.
Refinements of our present discussion will be feasible, once the
results of the susy analysis of the Higgs events by the LEP
experimental Collaborations are available.

\section{A supersymmetric interpretation of the LEP Higgs events}

In Ref. \cite{leplast} it is shown that the LEP Higgs events at
center-of-mass energy $\sqrt s$ above 206 GeV are compatible with the
SM predictions for a SM Higgs with a mass of about 115 GeV.  Here we
determine the supersymmetric configurations which, in the Minimal
Supersymmetric Extension of the Standard Model (MSSM), could provide
events of the same topologies at approximately the same rates as in
the standard model.

In the standard model the Higgs particle $H_0$ may be produced in $e^+
e^-$ collisions either by Higgs--strahlung: $e^+ e^- \rightarrow Z
H_0$, or by $WW$ fusion: $e^+ e^- \rightarrow \nu_e \bar{\nu}_e H_0$
\cite{cern}.  In the present paper we only consider events with $\bar
q q \bar b b$ final states, then the $WW$ fusion mechanism is not
considered here.

The main mechanisms for production of the neutral Higgs bosons: $h, A,
H$ of the MSSM at LEP2 are Higgs--strahlung: $e^+ e^- \rightarrow Z h$
(or $Z H$) and associated pair production: $e^+ e^- \rightarrow A h$
(or $A H$). Here $h$ and $H$ are the lighter and the heavier CP-even
Higgs boson, respectively, and $A$ is the CP-odd one.  The cross
sections for these processes are related to the SM cross--section for
Higgs--strahlung, $\sigma_{SM}$, by the formulae \cite{gh,dkz}

\begin{eqnarray}
& &\sigma (e^+  e^- \rightarrow Z h) = \sin^2(\alpha - \beta) \sigma_{SM}
\label{eq:Zh}\\
& &\sigma (e^+  e^- \rightarrow A h) =
\cos^2 (\alpha - \beta) \bar{\lambda}  \sigma_{SM},
\label{eq:Ah}
\end{eqnarray}

\noindent
where $\bar{\lambda} =
\lambda^{3/2}_{Ah}/[\lambda^{1/2}_{Zh}(12 {m_Z}^2/s +
\lambda_{Zh})]$ and $\lambda_{ij} = [1 - (m_i + m_j)^2/s][1 - (m_i -
m_j)^2/s]$. The cross sections for production of $H$ are derived from
those for production of $h$ (Eqs. (\ref{eq:Zh},\ref{eq:Ah})), by
interchanging $\sin^2(\alpha - \beta)$ with $\cos^2 (\alpha - \beta)$.

The CP--even  Higgs bosons are defined, in terms of the neutral
components of the original Higgs doublets, as

\begin{eqnarray}
& & H = \cos \alpha H_1^0 + \sin \alpha H_2^0 \\
& & h = - \sin \alpha H_1^0 + \cos \alpha H_2^0
\label{eq:higgs}
\end{eqnarray}

In the diagonalization of the mass matrix we impose the mass
hierarchy: $m_h < M_H$ and take the angle $\alpha$ in the range
$[-\frac{\pi}{2},+\frac{\pi}{2}]$. The angle $\beta$ is defined, as
usual, as the ratio of the two Higgs vev's: $\tan\beta =
<H^0_2>/<H^0_1>$.

We now consider the various possibilities for having, within the MSSM,
a final state $\bar q q \bar b b$ with the properties of the relevant
LEP Higgs events \cite{leplast}: we require that the particle
generating the pair $\bar q q$ has a mass of 91 GeV (this, in turn,
could be either a $Z$ in a Higgs--strahlung process, or an $h$, or an
$A$ in an associated pair production), and that the particle
generating the pair $\bar b b$ has an invariant mass of 115 GeV (this, in
turn, may be any of the three neutral Higgs bosons).  

Notice that, from LEP searches at lower center-of-mass energies
\cite{lep202}, an $h$ or an $A$ Higgs boson with a mass of 91 GeV is
already excluded for low values of $\tan\beta$, and is quite close to
the boundary of the allowed 90\% C.L. region for large values of
$\tan\beta$. We obviously take these limits into account in our
analysis. Therefore, the associated production topologies are possible
only for large $\tan\beta$, and also in this case they are borderline.
We choose to include them anyway, whenever they are possible.

To allow for experimental uncertainties in the reconstructed invariant
masses, a variation of $\pm$ 2 GeV is added to the $\bar b b$ Higgs
invariant mass of 115 GeV. For the same reason, we consider an
uncertainty also for the $\bar q q$ invariant mass associated to a
Higgs boson (i.e., in the associated production channels), but anyway
including the above mentioned lower limits on the susy Higgs
masses\cite{lep202}.

Then we can list the following independent categories of susy events
which can be considered to be compatible with the relevant LEP Higgs
events (other categories are not possible for the given topologies and
required mass assignments):

(1) $e^+ + e^- \rightarrow Z + h \rightarrow (\bar q,q) + (\bar b,b)$,
with the conditions that $m_h = 115 \pm 2$ GeV (the $\bar qq$ pair is
associated to the $Z$, while the $\bar bb$ pair is associated to the
$h$ boson);

(2) $e^+ + e^- \rightarrow h + A \rightarrow (\bar q,q) + (\bar b,b)$,
with the conditions that $m_h = (91\div 93)$ GeV and $m_A = 115 \pm 2$
GeV (the $\bar qq$ pair is associated to the $h$ boson, while the
$\bar bb$ pair is associated to the $A$ boson);

(3) $e^+ + e^- \rightarrow Z + H \rightarrow (\bar q,q) + (\bar b,b)$,
with the conditions that $m_H = 115 \pm 2$ GeV (the $\bar qq$ pair
is associated to the $Z$, while the $\bar bb$ pair is associated
to the $H$ boson);

(4) $e^+ + e^- \rightarrow Z + h, Z + H \rightarrow (\bar q,q) + (\bar
b,b)$, with the conditions that $m_h, m_H = 115 \pm 2$ GeV (the $\bar
qq$ pair is associated to the $Z$, while the $\bar bb$ pair is associated
either to the $h$ or the $H$ boson);

(5) $e^+ + e^- \rightarrow h + A, Z + H \rightarrow (\bar q,q) + (\bar
b,b)$, with the conditions that $m_A, m_H = 115 \pm 2$ GeV and $m_h =
(91\div 93)$ GeV (the $\bar qq$ pair is associated to the $h$ or the
$Z$, while the $\bar bb$ pair is associated either to the $A$ or to
the $H$ boson);

(6) $e^+ + e^- \rightarrow Z + H, A + H \rightarrow (\bar q,q) + (\bar
b,b)$, with the conditions that $m_A = (91\div 93)$ GeV  and $m_H = 115 \pm
2$ GeV (the $\bar qq$ pair is associated to the $Z$ or to the $A$,
while the $\bar bb$ pair is associated to the $H$ boson).

Since the LEP Higgs events are at the level of the SM predictions we
extract the compatible supersymmetric configurations by requiring that
the expected susy predictions are at this same level, within an
uncertainty of 20\%. For instance, in case (1) we impose that

\begin{equation}
0.8 \leq \, \sin^2(\alpha - \beta) 
\frac{BR_{\rm MSSM}(h \rightarrow \bar b b)}
{BR_{\rm SM}(H_0 \rightarrow \bar b b)} \, \leq \, 1.2 \, , 
\label{eq:h}
\end{equation}

\noindent
and in case (2) that

\begin{equation}
0.8 \, \leq \cos^2(\alpha - \beta) \bar{\lambda}
\frac{
BR_{MSSM}(h \rightarrow \bar q q) BR_{\rm MSSM}(A \rightarrow \bar b b)}
{BR_{\rm SM}(Z \rightarrow
\bar q q) BR_{\rm SM}(H_0 \rightarrow \bar b b)}
\, \leq \, 1.2 \, , 
\label{eq:H}
\end{equation}

\noindent
and similarly for the other cases. $BR_{SM}$ and $BR_{MSSM}$ denote the
branching ratios in the standard model and in MSSM, respectively.  

For each of the categories defined above and whenever necessary, we have
checked that production channels with Higgs mass assignements outside the two
ranges: $(91\div 93)$ GeV and $115 \pm 2$ GeV would not produce an exceedingly
large excess of events which have not actually been observed. Notice that, due
to lack of analysis by the experimental Collaborations within MSSM, our
analysis takes into account only a selection of events based on ranges of
masses, and not other selection criteria based for instance on kinematics. A
more complete analysis taking account also of the specific kinematical
constraints will be feasible, once the analysis within MSSM by the LEP
Collaborations are available.

We remind that the couplings of the bosons $h$, $H$ and $A$ to the up-type and
down--type quarks are proportional to $m_q k_u$ and $m_q k_d$, where $m_q$
denotes the quark mass, and the coefficients $k_u$, $k_d$ are given by

\begin{center}
\begin{tabular}{c|ccccc} 
~ & $h$ &~~ & $H$ &~~ & $A$ \\ \hline
$k_u$~~ & $\cos\alpha / \sin\beta$ &~~ & $\sin\alpha / \sin\beta$ &~~ & $1 /
\tan\beta$ \\
\\
$k_d$~~ & $-\sin\alpha / \cos\beta 
-\epsilon \cos (\alpha - \beta) \tan \beta$
&~~ & $\cos\alpha / \cos\beta-\epsilon \sin (\alpha - \beta) \tan
\beta$ &~~ 
& $\tan\beta (1+\epsilon)$
\end{tabular}
\end{center}
\noindent
In this table the entries include those radiative corrections which may be
sizeable at large $\tan\beta$. These corrections affect the couplings to
down--type quarks $k_d$, and are parametrized in terms of the quantity
$\epsilon\equiv 1/(1+\Delta)$, where $\Delta$ enters in the relationship
between the fermion running masses $m_d$ and the corresponding Yukawa couplings
$h_d$ \cite{qcd_corr}:
\begin{equation}
m_d=h_d <H_0^1>  (1+\Delta)
\label{eq:delta_d}
\end{equation}
These corrections take contributions mainly from gluino--squark,
chargino--squark and neutralino--stau loops \cite{qcd_corr}. Radiative
corrections to the Higgs-quark couplings $k_d$ affect the calculation of the
Higgs--decay branching ratios, of the neutralino--nucleus cross section and of
the neutralino cosmological relic abundance.

We notice that the correction to the relation between the b quark mass and its
Yukawa coupling defined in Eq.(\ref{eq:delta_d}) enters also in the
calculations of the $b \rightarrow s + \gamma$ decay \cite{bsg_carena}. For the
SUGRA model discussed below, it affects also the boundary conditions at the GUT
scale for the b Yukawa coupling \cite{sarid}. This in turn affects the
radiative symmetry breaking mechanism and the low--energy Higgs and sfermion
spectra \cite{bere1}. All these effects are included in our calculations.

To derive the specific supersymmetric configurations from the previous
conditions, one has to define the features of the susy scheme. In the
present paper we consider two models: a SUGRA model with unification
conditions at a grand unification scale (universal SUGRA) and an
effective model at the electroweak scale (effMSSM) \cite{our}.

The universal SUGRA model is parametrized in terms of five parameters: the
gaugino mass $m_{1/2}$, the scalar mass $m_0$, the trilinear coupling $A_0$,
$\tan \beta$, the sign of the Higgs--mixing coupling $\mu$.  $m_{1/2}$, $m_0$
and $A_0$ are defined at the unification scale.  In the present paper, these
parameters are varied in the following ranges: $50\;\mbox{GeV} \leq m_{1/2}
\leq 1\;\mbox{TeV},\; m_0 \leq 3\;\mbox{TeV},\; -3 \leq A_0 \leq +3,\; 1 \leq
\tan \beta \leq 50$.

The effMSSM model is given at the electroweak scale in terms of seven
independent parameters: the SU(2) gaugino mass $M_2, \mu, \tan\beta,
m_A$, a common mass for squarks $m_{\tilde q}$, a common mass for
sleptons $m_{\tilde l}$ and $A$; these parameters are varied in the
following ranges: $50\;\mbox{GeV} \leq M_2 \leq 1\;\mbox{TeV},\;
50\;\mbox{GeV} \leq |\mu| \leq 1\;\mbox{\rm TeV},\; 90\;\mbox{GeV}
\leq m_A \leq 1\;\mbox{TeV},\; 100\;\mbox{GeV} \leq m_{\tilde q},
m_{\tilde l} \leq 1\;\mbox{TeV},\; -3 \leq A \leq +3,\; 1 \leq \tan
\beta \leq 50$ ($m_A$ is the mass of the CP-odd neutral Higgs boson).
Our scanning of the susy parameter space, both in case of universal
SUGRA and of effMSSM takes also into account all available accelerator
constraints, including $b\to s+\gamma$ bounds. Further details about
our susy models and the ways in which the constraints are implemented
may be found in Refs.  \cite{our,noi}.

Let us turn now to the presentation of our results about the
supersymmetric configurations selected according to the criteria
explained above. These are provided by Figs. 1a, 1b in terms of the
angle $\alpha$ and of $\tan \beta$. In these figures and in all the
following ones, dots denote the representative points for events of
the category (1), i.e.  Higgs--strahlung of $h$, crosses denote events
of the category (3), i.e.  Higgs--strahlung of $H$, open dots denote
category (4) and finally filled dots denote category (6). We do not
find solutions for categories (2) and (5) above.

The main features displayed in the plots of Figs. 1a, 1b may be
understood in terms of the relations between the angles $\alpha$,
$\beta$ and $m_A$, these relations arising from the diagonalization of
the Higgs mass matrix. At values of $\tan\beta \sim 1$, $\sin(2\alpha)
\sim -1$ and $\alpha \sim -\pi/4 \sim \beta-\pi/2$. This implies that
$\sin^2(\alpha - \beta)\sim 1$ \cite{gbhs}.  Therefore, in this case, category (1)
events, i.e.  Higgs--strahlung of $h$ events, are able to reproduce
the LEP Higgs events with a production cross section at the level of
the SM one. On the contrary, when $\tan\beta$ is large,
$\sin(2\alpha)$ is small and $\cos(2\alpha) $ is usually close to 1,
except when $m_A$ is close to the mass of the $Z$ boson: in this
latter case, $\cos(2\alpha)$ can reach the values $1$ or $-1$,
depending on radiative correction terms.  Therefore, when $\tan\beta$
increases, $\alpha$ can cover the whole range $(-\pi/2,\pi/2)$ (the
sign depending on radiative correction terms in $\sin(2\alpha)$). When
$\alpha \sim 0 \sim \beta -\pi/2$ then $\sin^2(\alpha - \beta)\sim 1$
and a situation similar to the previous one holds: the LEP data can be
reproduced by Higgs--strahlung of $h$ events. On the other hand, when
$\alpha \sim \pi/2 \sim \beta$ or $\alpha \sim -\pi/2 \sim \beta
-\pi$, then $\cos^2(\alpha - \beta) \sim 1$: in this case the LEP data
can be reproduced by Higgs--strahlung of $H$ events.  Notice that in
order to have this last possibility, we need $m_A$ not too far from
$m_Z$. In the effMSSM scheme, this can be achieved easily, since $m_A$
is a free parameter.  On the contrary, in a SUGRA scheme, due to
radiative electroweak symmetry breaking, $m_A$ turns out to be a
decreasing function of $\tan\beta$ and it can be of the order of $m_Z$
only for $\tan\beta \gsim 40$ \cite{drees1,bere1}.

We wish to stress that the occurrence of the condition $\cos(2\alpha)
\sim -1$, and then $\cos^2(\alpha - \beta) \sim 1$, depends crucially
on the radiative corrections employed in the Higgs sector. In the
present paper we have used the results of Refs. \cite{radcorr}.

\section{Relic neutralinos: detection and cosmological abundance}

We turn now to the evaluation of the elastic neutralino--nucleon cross--section
$\sigma^{\rm (nucleon)}_{\rm scalar}$ and of the neutralino relic abundance
$\Omega_{\chi} h^2$ for the susy configurations selected on the basis of the
LEP Higgs events, and discussed in the previous Section.  The calculations of
$\sigma^{\rm (nucleon)}_{\rm scalar}$ have been performed with the formulae
reported in Refs. \cite{our,noi,noi6}; set 1 for the quantities
$m_{q}<\bar{q}q>$'s has been used (see Ref. \cite{noi6} for definitions; set 1
is on the conservative side of the range considered in \cite{noi6}); the
evaluation of $\Omega_{\chi} h^2$ follows the procedure given in \cite{noiom}.
  
Figs. 2a, 2b display the scatter plots for $\sigma_{\rm scalar}^{(\rm
  nucleon)}$ versus $\Omega_{\chi} h^2$ for effMSSM and universal SUGRA,
respectively.  For universal SUGRA only the results corresponding to positive
values of $\mu$ are displayed, since, for negative values, the constraint on $b
\rightarrow s + \gamma$ implies a large suppression of $\sigma_{\rm
  scalar}^{(\rm nucleon)}$.  The two horizontal lines bracket the sensitivity
region defined by Eq.  (\ref{eq:section}), when $\xi = 1$. The two vertical
lines denote a favorite range for the cosmological matter density $\Omega_{\rm
  m} h^2$: $0.05 \leq \Omega_{\rm m} h^2 \leq 0.3$, as derived from a host of
observational data. Notice that the most recent determinations of cosmological
parameters \cite{bmfr} appear to pin down the matter relic abundance to a
narrower range 0.08 $\lsim \Omega_{\rm m} h^2 \lsim$ 0.21.  However, some
caution in taking this range too rigidly is advisable, since some
determinations of cosmological parameters are still subject to fluctuations.
We point out that in the present paper we are not restricting ourselves to any
particular interval of $\Omega_{\rm m} h^2$.  Only some features of Figs. 3a,
3b depend on the actual value employed for the minimum amount of matter
necessary to reproduce the halo properties correctly.

Fig. 2a shows the quite remarkable result that in effMSSM almost all
susy configurations of the Higgs events of categories (3), (4), (6),
and a sizeable part of those of category (1) fall in the range of
detectability by current WIMP direct searches. Furthermore, part of
these configurations entail neutralinos of great cosmological
interest. This is also true for the SUGRA scheme of Fig. 2b.

It is interesting to examine the nature of the contributions which dominate the
scalar neutralino--nucleon cross--section. Let us do it for events of the
categories (1) and (3).  In the case of production of an $h$ by
Higgs--strahlung, one has $\sin^2(\alpha - \beta) \sim 1$, then $|\tan \alpha|
\sim 1/\tan\beta$, with the consequence that $|k_u| \sim 1, |k_d| \sim 1$ for
$h$, and $|k_u| \sim 1/\tan\beta$, $|k_d| \sim \tan \beta (1+\epsilon)$ for
$H$. From these properties and the fact that the coherent cross-section takes
its dominant contribution from the strange-quark content of the nucleon
\cite{noi6}, we derive that in this case $\sigma_{\rm scalar}^{(\rm nucleon)}$
is dominated (except for values of $\tan\beta$ close to 1) by exchange of an
$H$ boson, when $H$ is relatively light and $\epsilon$ is not close to $-1$.
However, the $H$ boson is not bounded from above in mass and can be naturally
heavy: in this case the $\sigma_{\rm scalar}^{(\rm nucleon)}$ for
Higgs-exchange can become small.  The situation is reversed in the case of
production by Higgs--strahlung of an $H$.  In fact, now one has $\cos^2(\alpha
- \beta) \sim 1$, then $|\tan \alpha| \sim \tan \beta$, which implies $|k_u|
\sim 1/\tan\beta, |k_d| \sim \tan \beta (1+\epsilon)$ for $h$ and $|k_u| \sim
1, |k_d| \sim 1$ for $H$.  Thus, in this case $\sigma_{\rm scalar}^{(\rm
  nucleon)}$ is dominated by exchange of an $h$, except when $\epsilon$ is
close to $-1$.  This boson is always light, and even more so here, since it has
to be lighter than $H$, whose mass is fixed to be about 115 GeV in order to
match the requirement for the LEP events; then $\sigma_{\rm scalar}^{(\rm
  nucleon)}$ is sizeable. It is interesting to notice that in this case, i.e.
Higgs--strahlung of an $H$ in the LEP data, where all the three susy Higgses
are close in mass, also the neutralino pair--annihilation cross--section
$\langle \sigma_{\rm ann} v \rangle$, which is responsible for the relic
abundance, is dominated by Higgs exchange, namely by $A$--exchange into
fermions, mostly $\bar b b$ pairs. Since the dominant couplings to neutralinos
and to fermions of both $h$ and $A$ are similar ($\cos^2(\alpha - \beta) \sim
1$ and large $\tan\beta$), it is easy to show that $\langle \sigma_{\rm ann} v
\rangle/\sigma_{\rm scalar}^{(\rm nucleon)}$ $\sim 1.7 \cdot 10^{2} (M/{\rm
  GeV})^2 r^2/(4 r^2-1)^2$, where $r=m_\chi/M$ and $M$ denotes a common mass
scale for the three Higgs masses.  For the neutralino masses of interest here,
$m_\chi \sim M \sim 100$ GeV, the relic abundance is therefore easily expressed
as a function of $\sigma_{\rm scalar}^{(\rm nucleon)}$ as: $\Omega_\chi h^2
\sim 4\cdot 10^{-10}/(\sigma_{\rm scalar}^{(\rm nucleon)}/{\rm nbarn})$. This
implies that relic abundances of the order of 0.1 are obtained with
$\sigma_{\rm scalar}^{(\rm nucleon)}$ of the order of a few $\times 10^{-9}$
nbarn, as it is shown by the cross symbols in Figs. 2a and 2b.

We turn now to a comparison of our results with specific experimental
measurements. To do this we plot in Figs. 3a, 3b the quantity $\xi
\sigma^{\rm (nucleon)}_{\rm scalar}$ versus $m_{\chi}$.  $\xi$
is taken to be $\xi = {\rm min}\{1, \Omega_{\chi} h^2/(\Omega_{\rm m}
h^2)_{\rm min}\}$, in order to have rescaling in the neutralino local
density, when $\Omega_{\chi} h^2$ turns out to be less than
$(\Omega_{\rm m} h^2)_{\rm min}$ (here $(\Omega_{\rm m} h^2)_{\rm
  min}$ is set to the value 0.05).

In Figs. 3a, 3b the solid line denotes the
frontier of the 3$\sigma$ annual--modulation region of Ref.
\cite{damalast}, when the uncertainties in $\rho_l$ and in the
dispersion velocity of a Maxwell--Boltzmann distribution, but not the
ones in other astrophysical quantities, are taken into account.
Effects due to a possible bulk rotation of the dark halo
\cite{dfs,noi5}, or to an asymmetry in the WIMP velocity distribution
\cite{vu,ecz,amg} would move this boundary towards higher values of
$m_{\chi}$.

We note that Figs. 3a, 3b show that the annual--modulation effect of
Ref. \cite{damalast} is quite compatible with supersymmetric
configurations involved in the LEP Higgs events.

\section{Conclusions}

Motivated by the intriguing results of the LEP Collaborations about a
hint for a possible neutral Higgs with a mass of about 115 GeV, we
have considered what might be the consequences for dark matter, in
case the LEP Higgs events are interpreted as due to supersymmetric
neutral Higgs bosons in a Minimal Supersymmetric Extension of the
Standard Model.

Using two extreme susy schemes, a universal SUGRA and an effective
scheme at the electroweak scale (effMSSM), we have proved that the
supersymmetric configurations extracted from the LEP data are
compatible with relic neutralinos of cosmological interest and of
relevance for the current WIMP direct searches. Quite remarkably, the
analyzed susy configurations would fit the annual--modulation effect
of Ref. \cite{damalast}. It is obvious that the same
conclusions apply for susy SUGRA schemes where some of the
unification conditions at the grand unification are partially relaxed
(non--universal SUGRA schemes).

Various cautionary comments are in order here. First, the effect seen
at LEP is only at a significance of 2.9 $\sigma$ , and no confirmation (or
disproof) of this most relevant subject will unfortunately be
available for quite a long time.  Secondly, a detailed analysis of the
current LEP data in terms of susy Higgs bosons has still to be
completed by the LEP Collaborations; once this is available, some
of the estimations performed in our Sect. II might be subject to
refinements.

The very nature of the LEP Higgs data necessarily confers to any
analysis of these results a somewhat speculative character. However,
due to the important properties at stake, it is very intriguing to
work out the various ensuing possible scenarios. Our analysis shows
that a quite coherent picture for supersymmetry and particle dark
matter may come out from the merging of quite independent
observations: measurements at accelerators and detection of relic
particles around us.

\section{Note Added}

After submission of our paper a new accurate experimental determination of the
muon anomalous magnetic moment appeared (H.N. Brown et al., hep-ex/0102017).
This data, if compared with theoretical evaluations in M. Davier and A.
H\"ocker, Phys. Lett. B435, 427 (1998), would show a deviation of 2.7 $\sigma$.
This has determined an outburst of theoretical papers where this possible
deviation is attributed to supersymmetry, and the relevant implications are
derived.  However, as pointed out in F.J. Yndur\'ain, hep-ph/0102312, other
standard--model evaluations of $a_{\mu}$ are in fair good agreement with the
experimental data. Thus, for the time being, it appears safer to use these data
as a constraint on supersymmetry, rather than a sign of it. Employing the set
of theoretical results reported in F.J. Yndur\'ain, hep-ph/0102312, we find
that the contribution of supersymmetry to the anomalous moment is constrained
by $-600 \leq a_{\mu}^{susy} \cdot 10^{11} \leq 800$. This constraint has been
implemented in our scanning of the supersymmetric parameter space.

\acknowledgements 

We thank the referee for useful comments on the implications of the susy QCD
corrections in higgs-quark couplings for the neutralino--nucleon cross section.
This work was partially supported by the Research Grants of the Italian
Ministero dell'Universit\`a e della Ricerca Scientifica e Tecnologica (MURST)
within the {\sl Astroparticle Physics Project}.

\newpage
\begin{figure}[t]
  \hbox{
    \psfig{figure=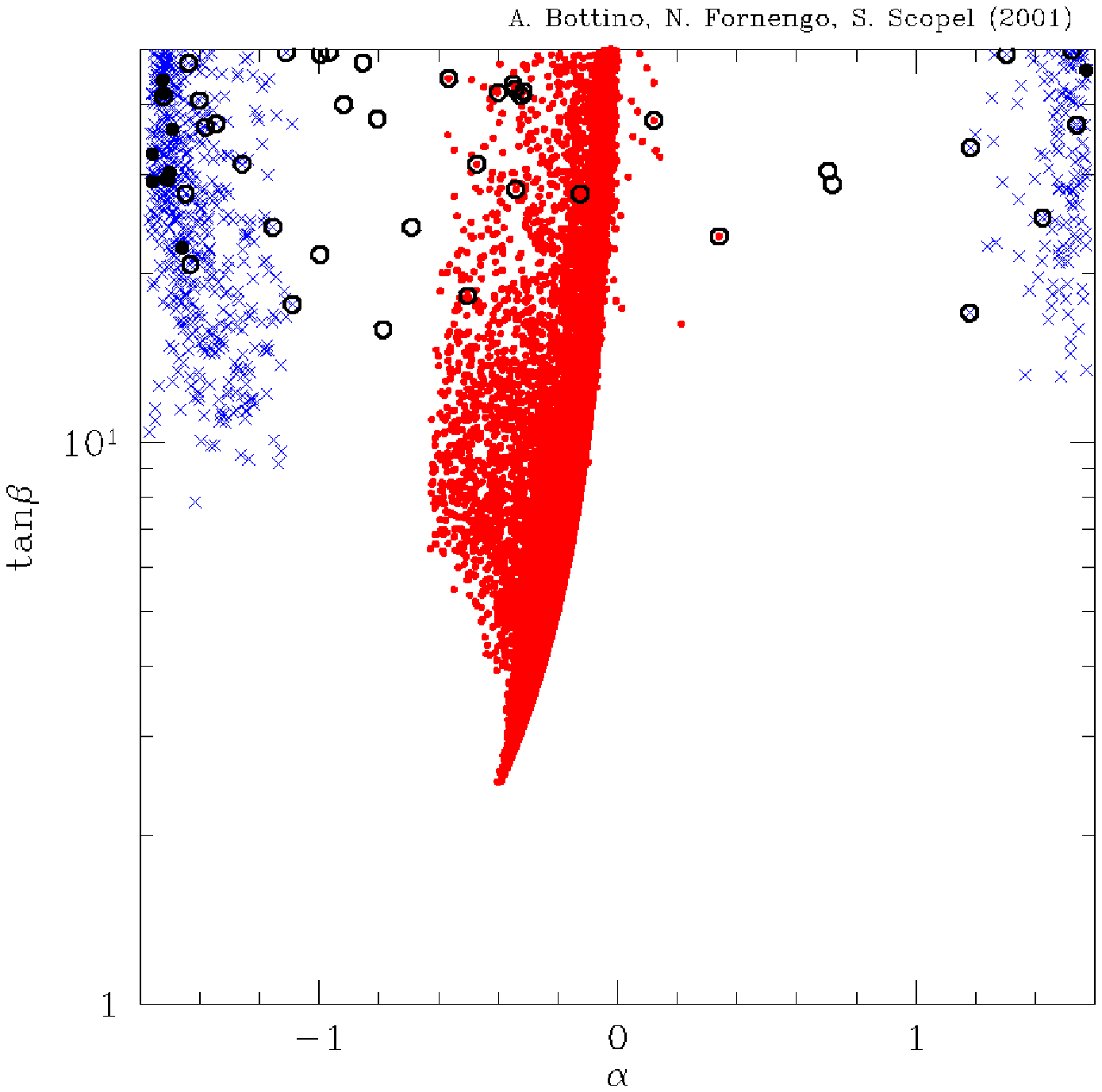,width=8.2in,bbllx=40bp,bblly=160bp,bburx=700bp,bbury=660bp,clip=}
    } 
  
  { FIG.1a - Supersymmetric configurations selected according to the
    criteria for reproducing the relevant LEP Higgs events in the
    effMSSM, shown in the plane $\tan\beta$ vs. $\alpha$.
    Different points refer to different categories of events: dots
    refer to $e^+ + e^- \rightarrow Z + h \rightarrow (\bar q,q) +
    (\bar b,b)$, crosses refer to $e^+ + e^- \rightarrow Z + H
    \rightarrow (\bar q,q) + (\bar b,b)$, open dots refer to $e^+ +
    e^- \rightarrow Z + h, Z + H \rightarrow (\bar q,q) + (\bar b,b)$
    and finally filled dots refer to $e^+ + e^- \rightarrow Z + H, A +
    H \rightarrow (\bar q,q) + (\bar b,b)$.}
\end{figure}

\newpage
\begin{figure}[t]
\hbox{
\psfig{figure=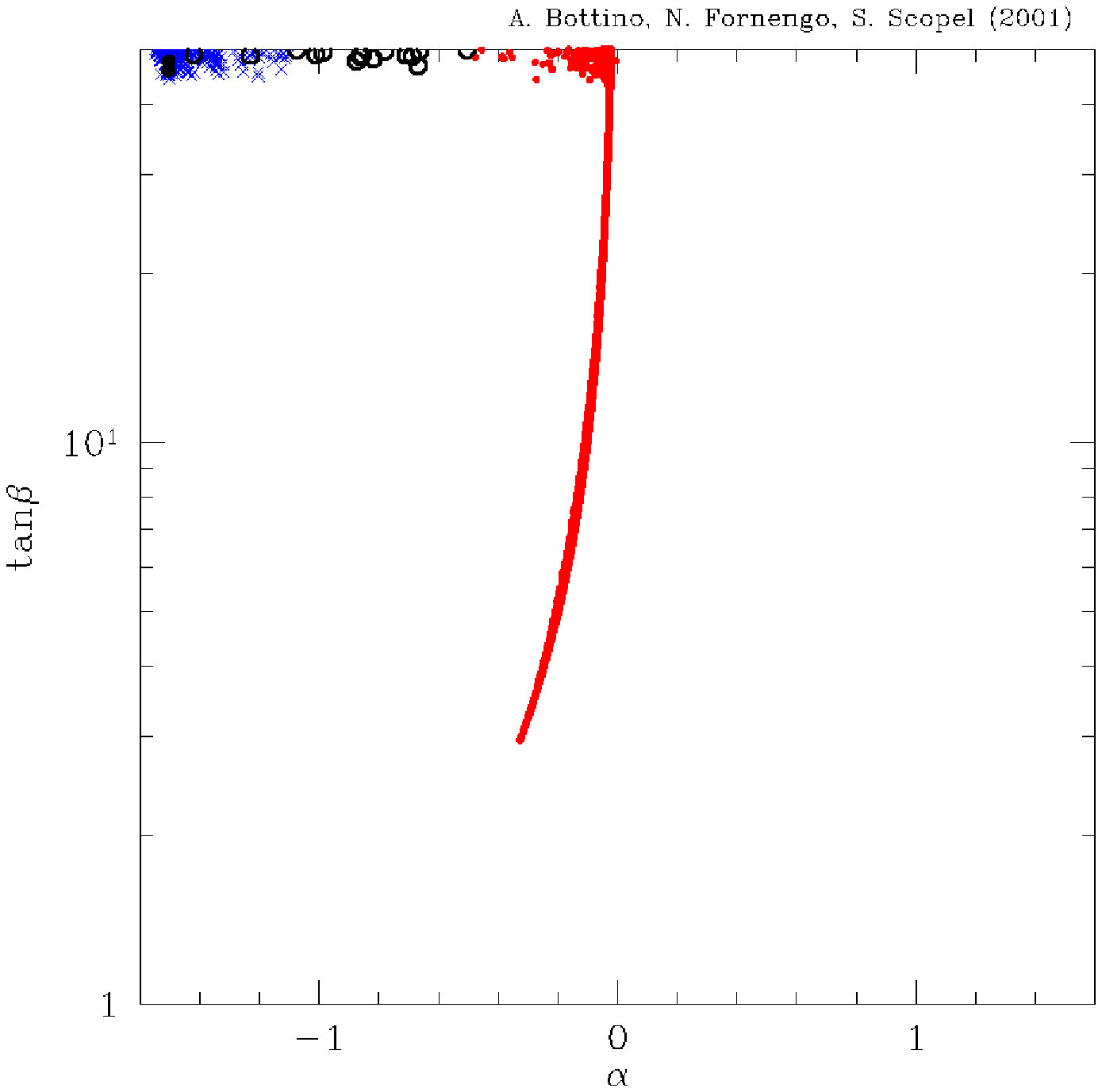,width=8.2in,bbllx=40bp,bblly=160bp,bburx=700bp,bbury=660bp,clip=}
  }

{ FIG.1b - Supersymmetric configurations selected according to the
  criteria for reproducing the relevant LEP Higgs events in universal
  SUGRA. Notations as in Fig.1a.  }
\end{figure}

\newpage
\begin{figure}[t]
\hbox{
\psfig{figure=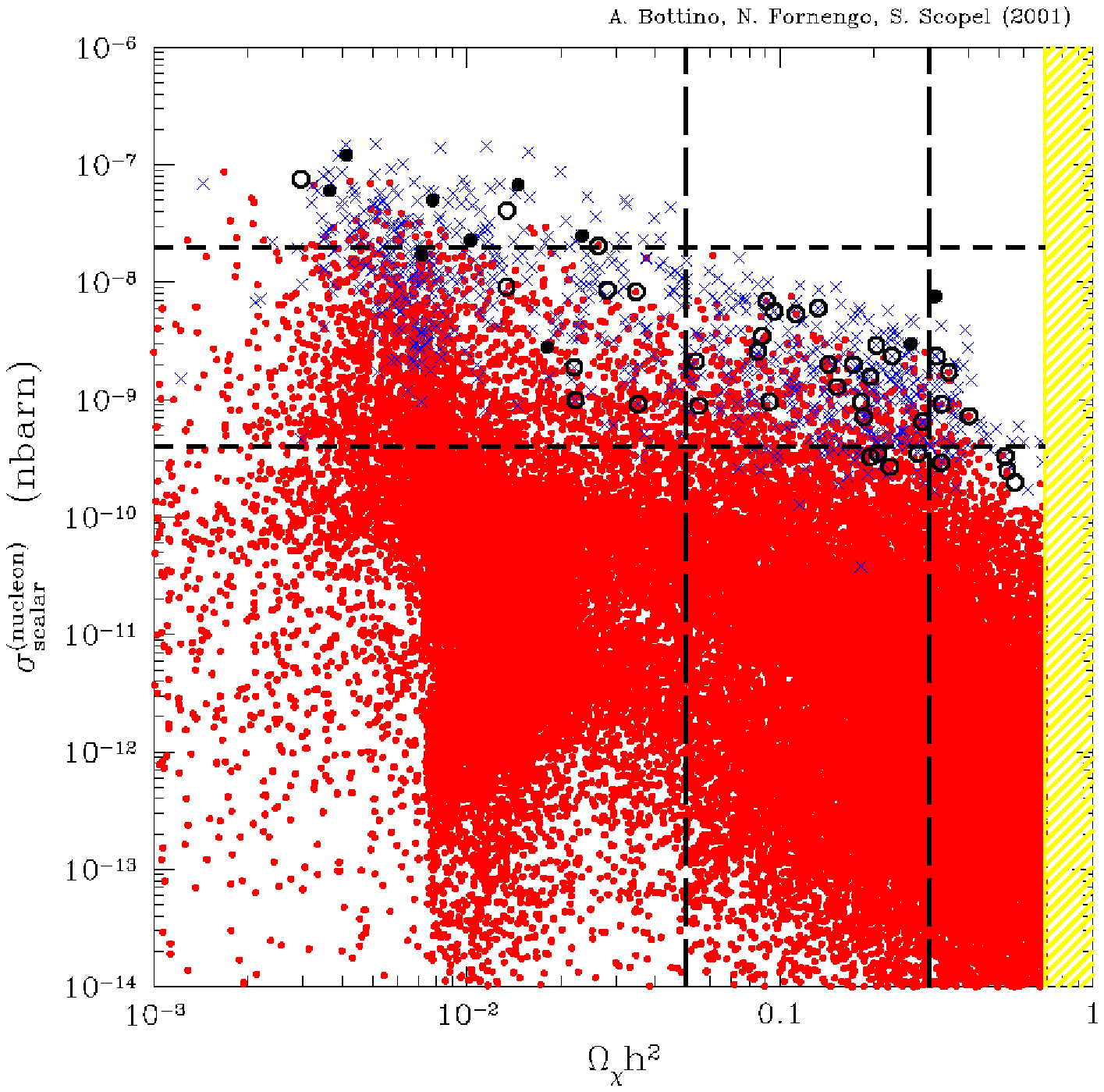,width=8.2in,bbllx=40bp,bblly=160bp,bburx=700bp,bbury=660bp,clip=}
  } 

{ FIG.2a - Scatter plot of the neutralino--nucleon scalar cross
  section $\sigma_{\rm scalar}^{(\rm nucleon)}$ versus the neutralino
  relic abundance $\Omega_{\chi} h^2$ for the effMSSM. Set 1 for the
  quantities $m_{q}<\bar{q}q>$'s is employed in the calculation of
  $\sigma_{\rm scalar}^{(\rm nucleon)}$. The two horizontal lines
  bracket the sensitivity region defined by Eq.  (\ref{eq:section}).
  The two vertical lines denote the range $0.05 \leq \Omega_{\rm m}
  h^2 \leq 0.3$.  The region where $\Omega_{\chi} h^2 > 0.7$ is
  excluded by current limits on the age of the Universe. Different
  points (notations as in Fig.1a) refer to different categories of
  events able to reproduce the relevant LEP Higgs events. }
\end{figure}

\newpage
\begin{figure}[t]
\hbox{
\psfig{figure=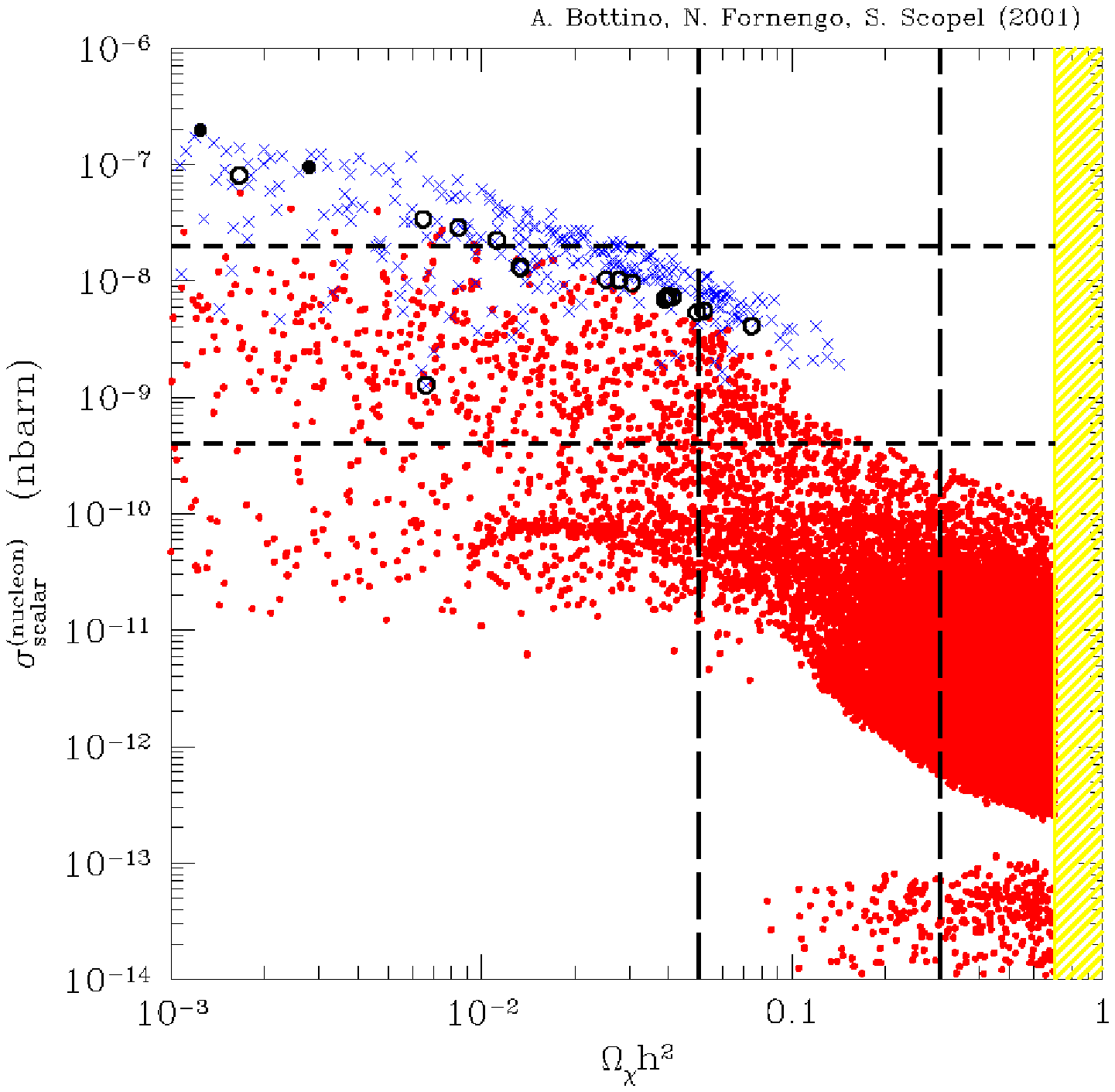,width=8.2in,bbllx=40bp,bblly=160bp,bburx=700bp,bbury=660bp,clip=}
  } 

{ FIG.2b - Scatter plot of the neutralino--nucleon scalar cross
  section $\sigma_{\rm scalar}^{(\rm nucleon)}$ versus the neutralino
  relic abundance $\Omega_{\chi} h^2$ for universal SUGRA. Notations
  and definitions as in Fig.2a.

}
\end{figure}

\newpage
\begin{figure}[t]
\hbox{
\psfig{figure=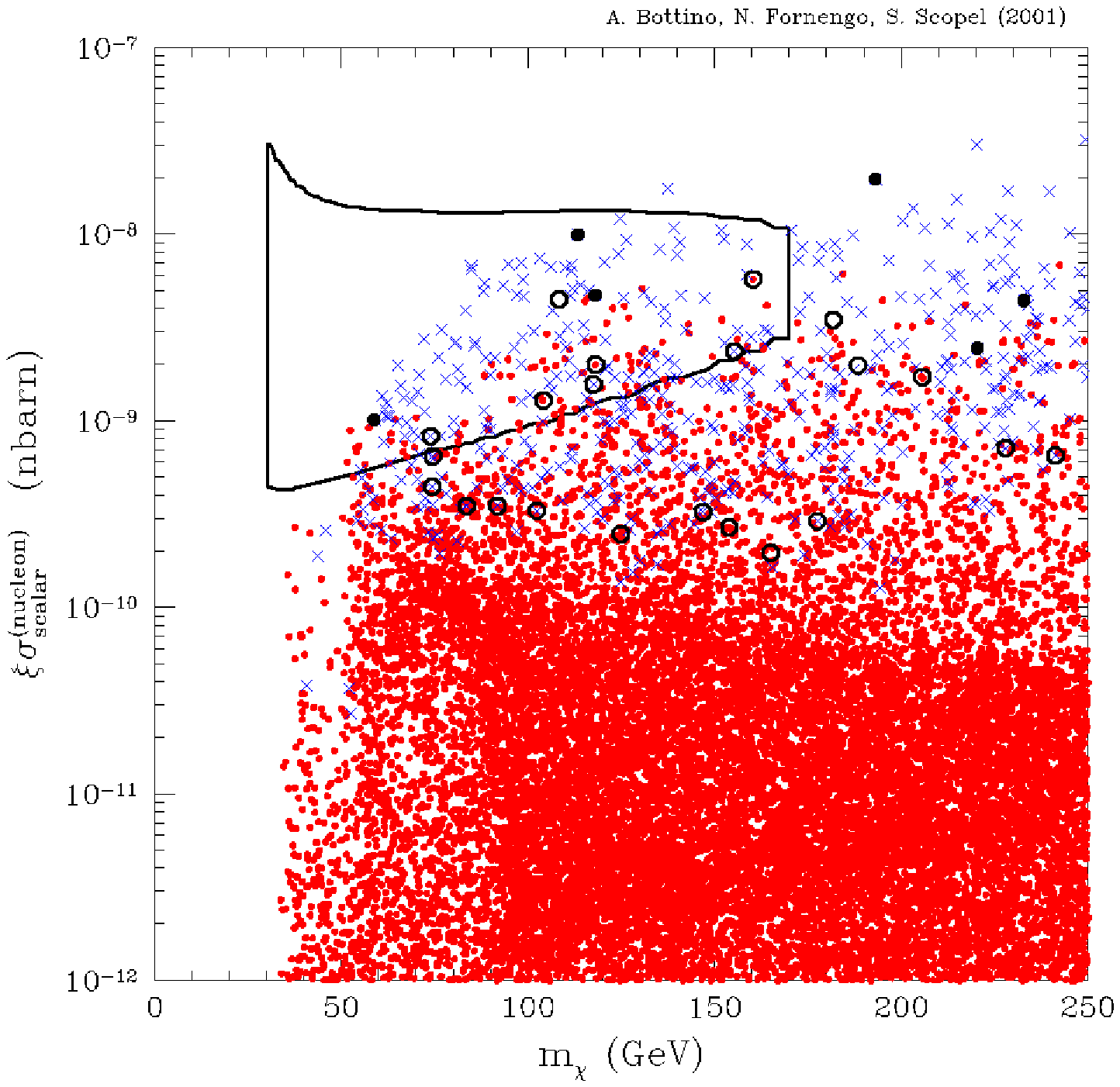,width=8.2in,bbllx=40bp,bblly=160bp,bburx=700bp,bbury=660bp,clip=}
  } 

{ FIG.3a - Scatter plot of $\xi \sigma_{\rm scalar}^{(\rm nucleon)}$
  versus the neutralino mass $m_{\chi}$ for the effMSSM. Set 1 for the
  quantities $m_{q}<\bar{q}q>$'s is employed in the calculation of
  $\sigma_{\rm scalar}^{(\rm nucleon)}$. Different points (notations
  as in Fig.1a) refer to different categories of events able to
  reproduce the relevant LEP Higgs events. The solid contour denotes
  the 3$\sigma$ annual--modulation region of Ref. \cite{damalast} when
  taking into account the uncertainties in the local dark matter
  density and in the dispersion velocity of the velocity distribution
  function of WIMPs in the galactic halo.  }
\end{figure}

\newpage
\begin{figure}[t]
\hbox{
\psfig{figure=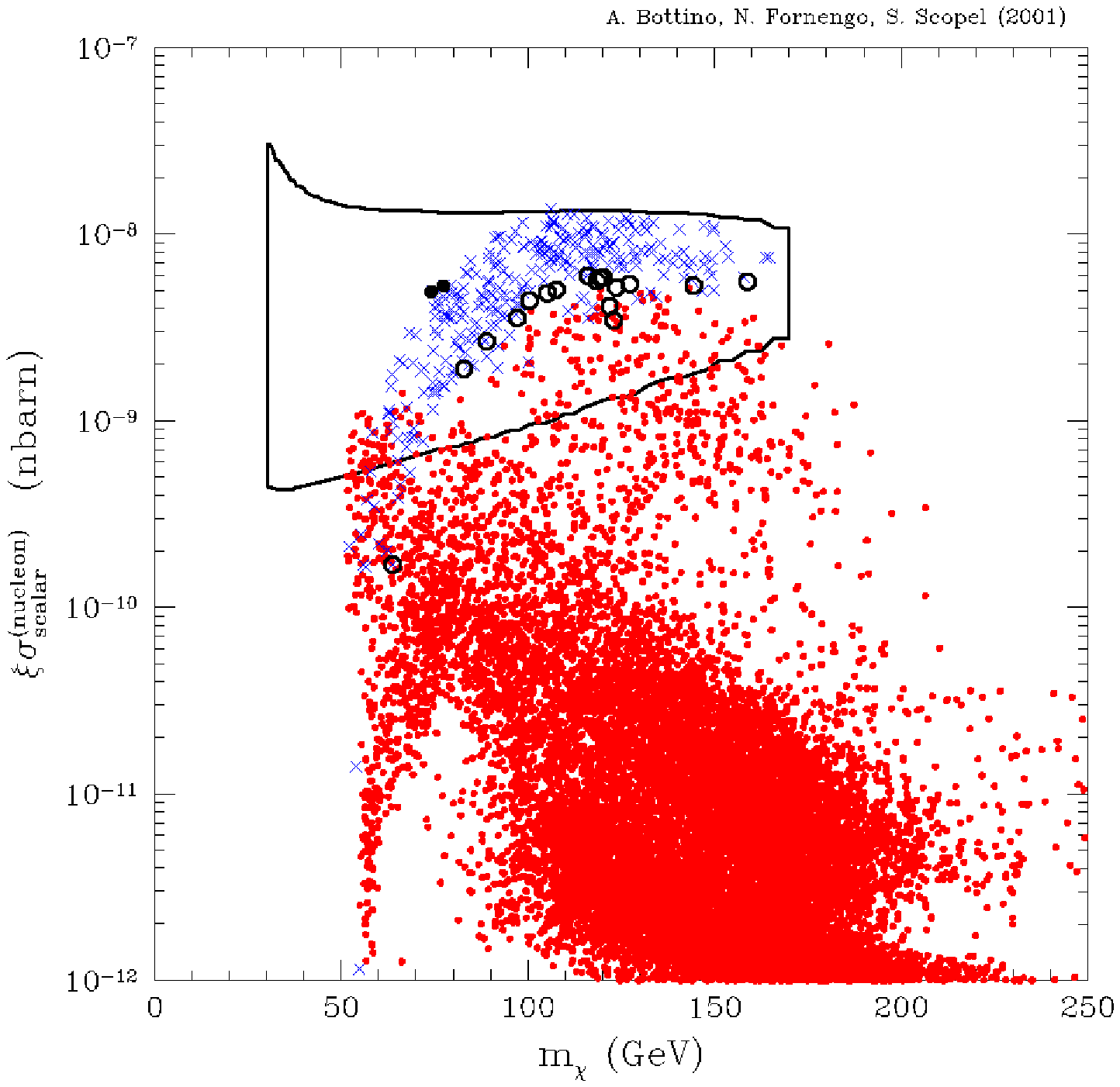,width=8.2in,bbllx=40bp,bblly=160bp,bburx=700bp,bbury=660bp,clip=}
}

{ FIG.3b - Scatter plot of $\xi \sigma_{\rm scalar}^{(\rm nucleon)}$
  versus the neutralino mass $m_{\chi}$ for universal SUGRA. Notations
  and definitions as in Fig.3a. }
\end{figure}

\end{document}